\documentclass[
 superscriptaddress,
 amsmath,amssymb,
 aps,
 floatfix
]{revtex4-2}

\usepackage{amsmath}
\usepackage{amssymb}
\usepackage{amsthm}
\usepackage[export]{adjustbox}
\usepackage{bm}
\usepackage{graphicx}
\usepackage{subfig}
\usepackage{hyperref}
\usepackage[capitalise]{cleveref}
\usepackage[a-1b]{pdfx}
\usepackage{algcompatible}

\newcommand{\mean}[1]{\left\langle#1\right\rangle}

\begin{document}

\title{Stability of Dining Clubs in the Kolkata Paise Problem with and without Cheating}

\author{Akshat Harlalka}
\email{akshatharlalka.ah@icloud.com}
\affiliation{Department of Computer Science, 
            Penn State University, 
            University Park, PA 16802}

\author{Andrew Belmonte}
\email{belmonte@psu.edu}
\affiliation{
            Department of Mathematics \& Huck Institute of Life Sciences,
            Penn State University, 
            University Park, PA 16802}

\author{Christopher Griffin}
\email{griffinch@psu.edu}
\affiliation{
            Applied Research Laboratory,
            Penn State University, 
            University Park, PA 16802}

\begin{abstract}
We introduce the idea of a dining club to the Kolkata Paise Restaurant Problem. In this problem, $N$ agents choose (randomly) among $N$ restaurants, but if multiple agents choose the same restaurant, only one will eat. Agents in the dining club will coordinate their restaurant choice to avoid choice collision and increase their probability of eating. We model the problem of deciding whether to join the dining club as an evolutionary game and show that the strategy of joining the dining club is evolutionarily stable. We then introduce an optimized member tax to those individuals in the dining club, which is used to provide a safety net for those group members who don't eat because of collision with a non-dining club member. When non-dining club members are allowed to cheat and share communal food within the dining club, we show that a new unstable fixed point emerges in the dynamics. A bifurcation analysis is performed in this case. To conclude our theoretical study, we then introduce evolutionary dynamics for the cheater population and study these dynamics. Numerical experiments illustrate the behaviour of the system with more than one dining club and show several potential areas for future research.
\end{abstract}

\maketitle

\section{Introduction}

The Kolkata Paise Restaurant Problem (KPRP) was first introduced in 2007 \cite{CMC07} during work on the Kolkata Paise Hotel Problem. Since then, it has been studied extensively \cite{BGCN12,CCGM17,CG19,GC17,DSC11,BMM13,BM21,SC20,GDCM12,CCCM15,CRS22,KPA22,CMC07,MK17,R13,Y10,GCCC14} in the econophysics literature.  In its simplest form, we assume $N \gg 1$ agents will choose among $N$ restaurants. Choice is governed by a distribution determined by an implicit ranking of the restaurants. The ranking represents the payoff of eating at a given restaurant. If two or more agents select the same restaurant, then the restaurant randomly chooses which agent to serve.

A broad overview of KPRP can be found in \cite{CCGM17,BMM13,CCCM15}. When all restaurants are ranked equally (i.e., have payoff $1$) and agents choose a restaurant at random, the expected payoff to each agent is easily seen to be approach $1-1/e$ as $N \to \infty$. Using stochastic strategies and resource utilization models, the mean payoff can be increased to $ \sim 0.8$ \cite{GCMC10}. Identifying strategies to improve on the uncoordinated outcome is a central problem in KPRP.

KPRP is an example of an anti-coordination game (such as Hawk-Dove) \cite{M12}. Other examples of this class of game are minority games \cite{CZ98,HZDH12} and the El Farol bar problem \cite{arthur1994, decara1999, FGH02, challet2004}. These types of games also emerge in models of channel sharing in communications systems \cite{AFGJ13,AFGJ13a,GK14}.

Learning in KPRP is considered in \cite{CRS22,GCMC10,GSC10} with both classical and quantum learning considered in \cite{CRS22}. Quantum versions of the problem are considered in \cite{CRS22,R13,Y10} and its relevance to other areas of physical modelling are considered in \cite{BM21,GCCC14,MK17,GDCM12} with phase transitions considered recently in \cite{BGCN12,SC20}. Distributed and coordinated solutions to optimizing agent payoff are discussed in \cite{CG19,GC17,DSC11,KPA22}.

In this paper, we use evolutionary game theory to study a group formation problem within the context of KPRP. We assume that some subset of the population of $N$ individuals forms a dining club. Individuals in the dining club coordinate their actions and will choose distinct restaurants from each other, thus increasing the odds that any individual within the dining club will eat. In this context, we show the following results:
\begin{enumerate}
    \item When all restaurants are ranked equally, membership in the dining club is globally stable. That is, asymptotically all players join the dining club (in the limit as $N \to \infty$). 
    \item When the dining club taxes its members by collecting food for redistribution to those members who did not eat, there is an optimal tax rate that ensures all members are equally well-fed. 
    \item When non-club members can choose to deceptively share in the communal food (freeload) of the dining club, a new unstable fixed point emerges. The fixed point corresponding to a population where all members join the dining club remains stable, but is no longer globally stable. We characterize the basin of attraction in this case. This effectively introduces a public goods game into the KPRP.
    \item We then use numerical analysis to study the case where two dining clubs are active. We numerically illustrate the existence of equilibrium surfaces where multiple dining clubs can exist simultaneously along with non-group members as a result of group taxation (food sharing), cheating (freeloading), and cheating detection. 
\end{enumerate}

The remainder of this paper is organized as follows: In \cref{sec:Math}, we analyse an evolutionary model of KPRP with a dining club. We study resource distribution through taxation and cheating in \cref{sec:Cheating}. Cheating is modelled in an evolutionary context in \cref{sec:EvolveCheating}. KPRP with multiple dynamic clubs is studied numerically in \cref{sec:MultipleClubs}. Finally, in \cref{sec:Conclusions} we present conclusions and future directions.

\section{Mathematical Analysis}\label{sec:Math}
We first study KPRP with a single dining club. Let $g$ be the size of the dining club and let $n$ be the size of the free population with total population given by $N = g+n$. The probability that an individual in the dining club eats is given by
\begin{equation*}
p_g(n, g) = \sum_{k=0}^n \binom{n}{k}\left(\frac{n+g-1}{n+g}\right)^{n-k}\left(\frac{1}{n+g}\right)^k\frac{1}{k+1},
\end{equation*}
while the probability that a free individual eats is given by
\begin{equation*}
p_n(n,g) =
\sum_{k=0}^{n-1} \binom{n-1}{k}\left(\frac{n}{n+g}\frac{1}{k+1} + 
\frac{g}{n+g}\frac{1}{k+2} \right)\left(\frac{n+g-1}{n+g}\right)^{n-k-1}\left(\frac{1}{n+g}\right)^k.
\end{equation*}
If we assume  $g = \alpha n$ and sum over $k$, then we can rewrite $p_g(n,g)$ in closed form as
\begin{equation*}
p_g(n, \alpha) = \frac{\left(1-\frac{1}{\alpha  n+n}\right)^n \left((\alpha +1) n
   \left(\left(\frac{1}{\alpha  n+n-1}+1\right)^n-1\right)+1\right)}{n+1}.
\end{equation*}
Likewise, $p_n(n,g)$ can be written as
\begin{equation*}
p_n(n,\alpha) = \frac{\left(1-\frac{1}{\alpha  n+n}\right)^n}{n+1}
\left\{ \alpha ^2 n  +\alpha 
   n -\alpha -n-1
-\left[(\alpha
   +1) ((\alpha -1) n-1) \left(\frac{1}{\alpha  n+n-1}+1\right)^n\right]\right\}.
\end{equation*}
If we compute the limit as $n \to \infty$, this yields the asymptotic probabilities
\begin{equation}
p_g(\alpha) = \lim_{n \to \infty} p_g(n,\alpha) = \left(1-e^{-\frac{1}{\alpha +1}}\right) (\alpha +1),
\label{eqn:pg}
\end{equation}
and
\begin{equation}
p_n(\alpha) = \lim_{n \to \infty} p_n(n,\alpha) = -\alpha ^2+e^{-\frac{1}{\alpha +1}} \left(\alpha ^2+\alpha -1\right)+1.
\label{eqn:pn}
\end{equation}
For the remainder of this section and the next, we assume an infinite population. While it was easier to work with $g = \alpha n$ for the previous computation, for further analysis it is simpler to express $g$ as a fraction of the total population. Let
\begin{equation*}
\beta = \frac{g}{n+g} = \frac{\alpha}{1+\alpha}. 
\end{equation*}
Substituting
\begin{equation}
\alpha = \frac{\beta}{1-\beta}. 
\label{eqn:alpha}
\end{equation}
into \cref{eqn:pg,eqn:pn} yields the simplified forms,
\begin{align*}
p_g(\beta) &= \frac{1-e^{\beta -1}}{1-\beta}\quad \text{and}\\
p_n(\beta) &= \frac{-2 e \beta -e^{\beta } ((\beta -3) \beta +1)+e}{e (\beta -1)^2}.
\end{align*}
A simple plot shows that $p_g(\beta) \geq p_n(\beta)$ for all $\beta\in[0,1]$.
\begin{figure}[htbp]
\centering
\includegraphics[width=0.5\textwidth]{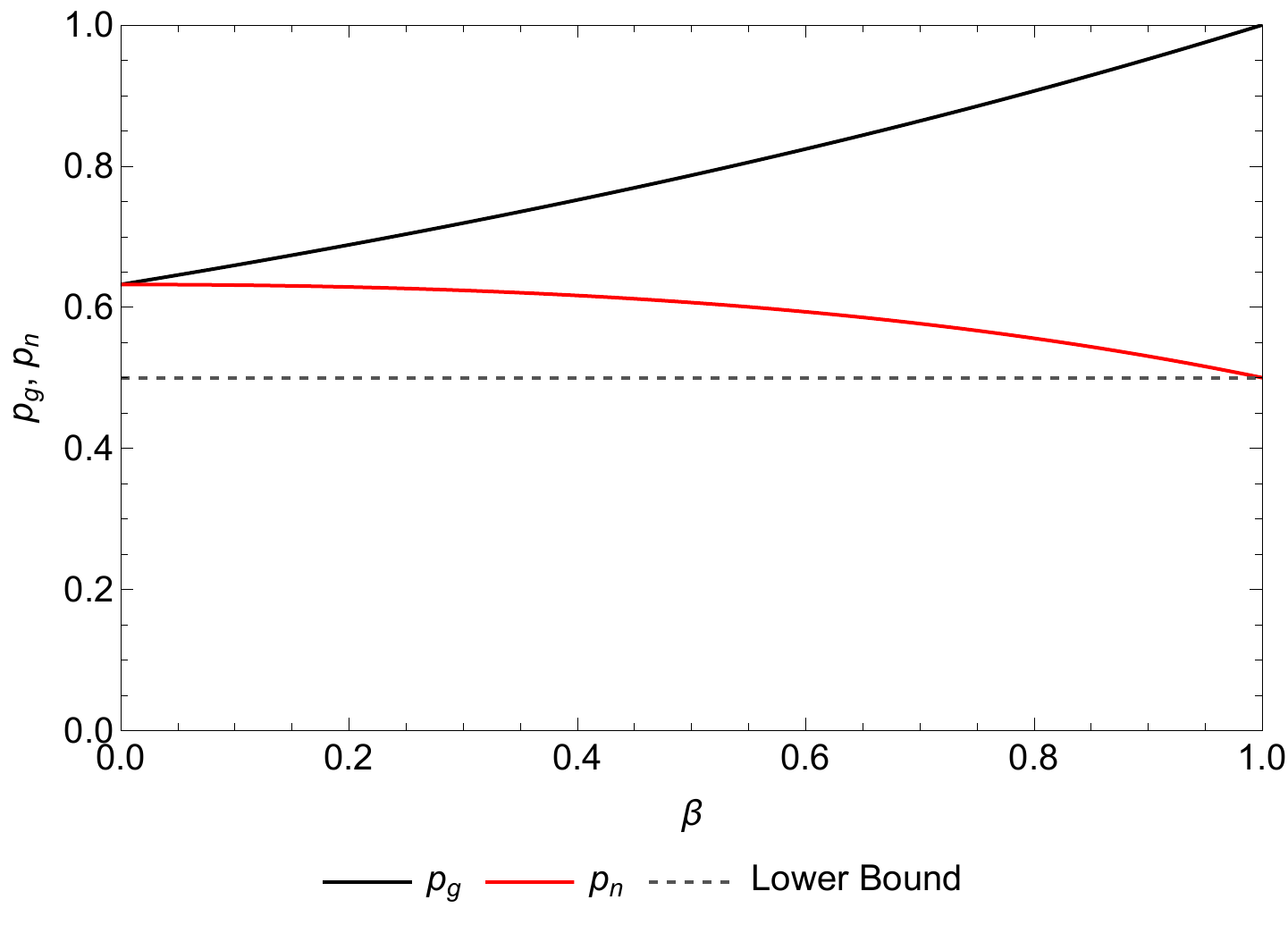}
\caption{A plot of $p_g(\beta)$ and $p_n(\beta)$ shows that it is always better for an individual to join the dining club than to remain independent.}
\label{fig:pnpg}
\end{figure}

\begin{figure}
\centering
\includegraphics[width=0.45\textwidth]{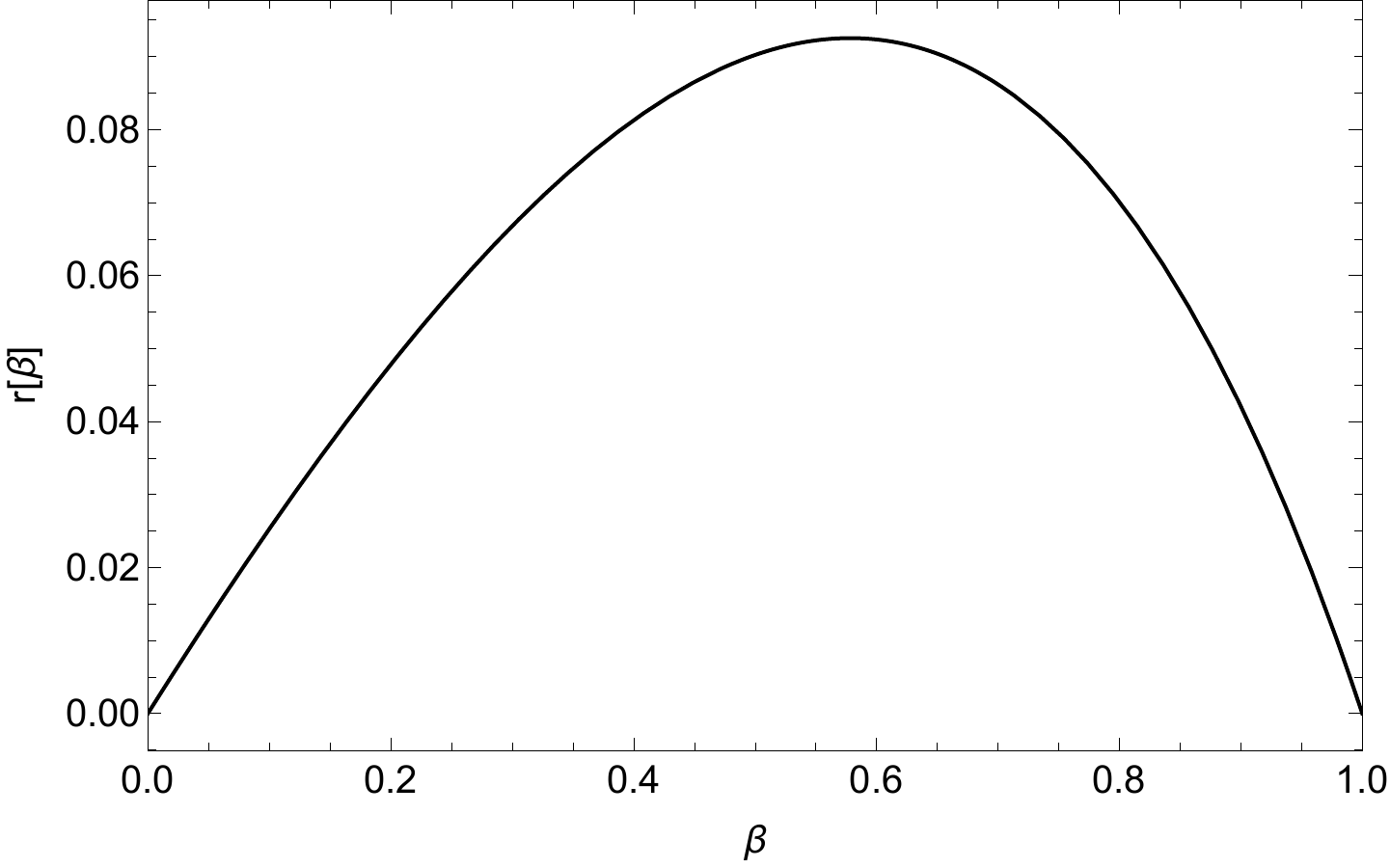}\quad
\includegraphics[width=0.45\textwidth]{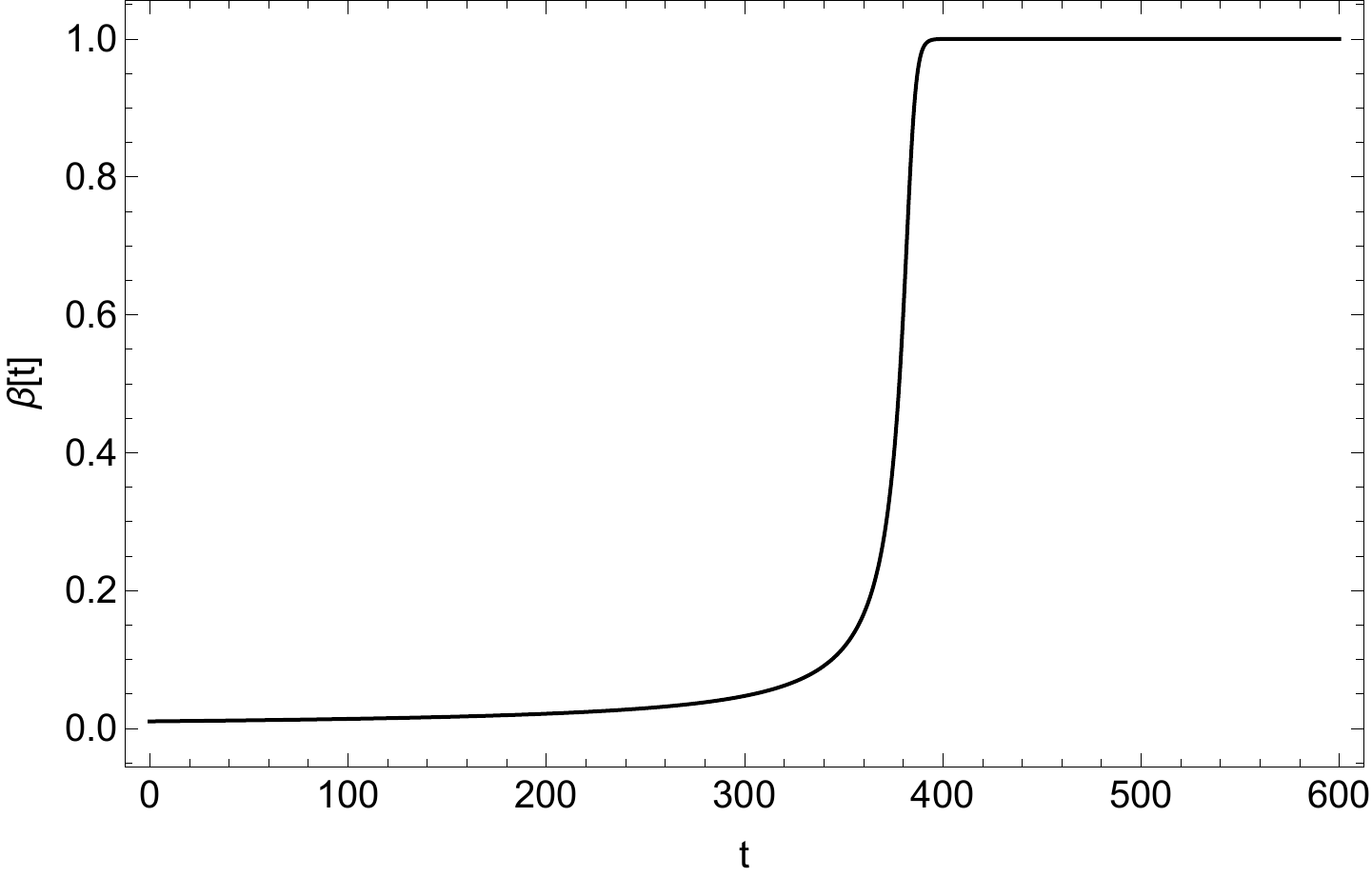}\quad
\caption{(Left) The growth rate of $r(t)$ is an unimodal positive function with zeros at $\beta = 0$ and $\beta = 1$. (Right) The solution curve for $\beta(t)$ assuming $\beta(0) = 0.01$.}
\label{fig:beta}
\end{figure}

Let $S_g$ be a random variable denoting the meal size for an individual in the dining club, and let $S$ be a random variable denoting the meal size for a randomly chosen member of the population. Then the probability of eating $p_g(\beta)$  is now easily seen as the expected meal size $\mean{S_g}$, with a meal size of $1$ corresponding to eating and a meal size of $0$ corresponding to not eating. Using this interpretation, and equating meal size with fitness, we assume the rate of growth of the dining club is given by
\begin{equation*}
\dot{g} = g\mean{S_g} = g p_g(\beta).
\end{equation*}
From \cite{GB17}, it follows that the proportion $\beta$ must follow the replicator dynamic
\begin{equation}
\dot{\beta} = \beta \left[p_g(\beta) - \bar{p}(\beta)\right] = \beta \left(\mean{S_g} - \mean{S}\right).
\label{eqn:BetaReplicator}
\end{equation}
The population mean $\bar{p}(\beta) = \mean{S}$ can be computed as
\begin{equation*}
\bar{p}(\beta) = \mean{S} = \frac{\alpha p_g(\alpha) + p_n(\alpha)}{1+\alpha},
\end{equation*}
and converted to an expression in $\beta$ using \cref{eqn:pg,eqn:pn,eqn:alpha} as,
\begin{equation*}
\mean{S} = \bar{p}(\beta) = e^{\beta -1} (\beta -1)+1.
\end{equation*}
Let 
\begin{equation*}
r(\beta) = p_g(\beta) - \bar{p}(\beta) = \mean{S_g} - \mean{S} 
= \frac{1-e^{\beta -1}}{1-\beta}
- \left(e^{\beta -1} (\beta -1)+1 \right),
\end{equation*}
be the growth rate of $\beta$. Then $r(0) = 0$ and we see that
$\lim_{\beta \to 1} r(\beta) = 0$. That is, \cref{eqn:BetaReplicator} has two fixed points. From \cref{fig:pnpg}, we must have $r(\beta) > 0$ for $0 < \beta < 1$. This is illustrated in \cref{fig:beta}  (left). It follows that $\beta(t)$ is described by a non-logistic sigmoid, as shown in \cref{fig:beta} (right).
We conclude that the decision to join the dining club is an evolutionarily stable strategy and the fixed point $\beta = 1$ is globally asymptotically stable while the fixed point $\beta = 0$ is asymptotically unstable.

\section{Social Safety Nets and Deceptive Free Loading}\label{sec:Cheating}
Suppose the dining club imposes a \textit{food tax} on its members at the rate $\kappa \in [0,1]$ so that if a diner is successful in obtaining food, then he reserves $\kappa \times 100\%$ of his meal to be shared with club members who choose a restaurant that is occupied by an independent individual. If we assume these resources are pooled and then shared equally, the expected meal size (normalized to the interval $[0,1]$) available for a club member who cannot obtain food on his own is given by
\begin{equation}
\tilde{p}_g(\beta) = \frac{gp_g(\beta)\kappa}{g - gp_g(\beta)} = \frac{p_g(\beta)\kappa}{1 - p_g(\beta)}.
\label{eqn:tildepg}
\end{equation}
Note that sharing (for any value of $\kappa$) does not affect the expected meal size obtained by a group member, since we have the expected meal size
\begin{equation}
\mean{S_g} = (1-\kappa) p_g(\beta) + [1-p_g(\beta)]\frac{p_g(\beta)\kappa}{1 - p_g(\beta)} = p_g(\beta).
\label{eqn:ES}
\end{equation}
We can construct a tax-rate that depends on $\beta$ and ensures all participants in the dining club receive the same meal size. Setting $\tilde{p}_g(\beta) = 1 - \kappa$ and solving, we obtain:
\begin{equation}
\kappa^* = 1 - p_g(\beta).
\label{eqn:OptimalKappa}
\end{equation}
Thus, as $\beta$ increases, the tax decreases. As a result of \cref{eqn:ES}, the right-hand-side of \cref{eqn:BetaReplicator} remains unchanged and the decision to join the dining club is still evolutionarily stable, even in the presence of sharing. That is $\beta = 1$ is still globally asymptotically stable. 

\begin{figure}[htbp]
\centering
\includegraphics[width=0.45\textwidth,valign=t]{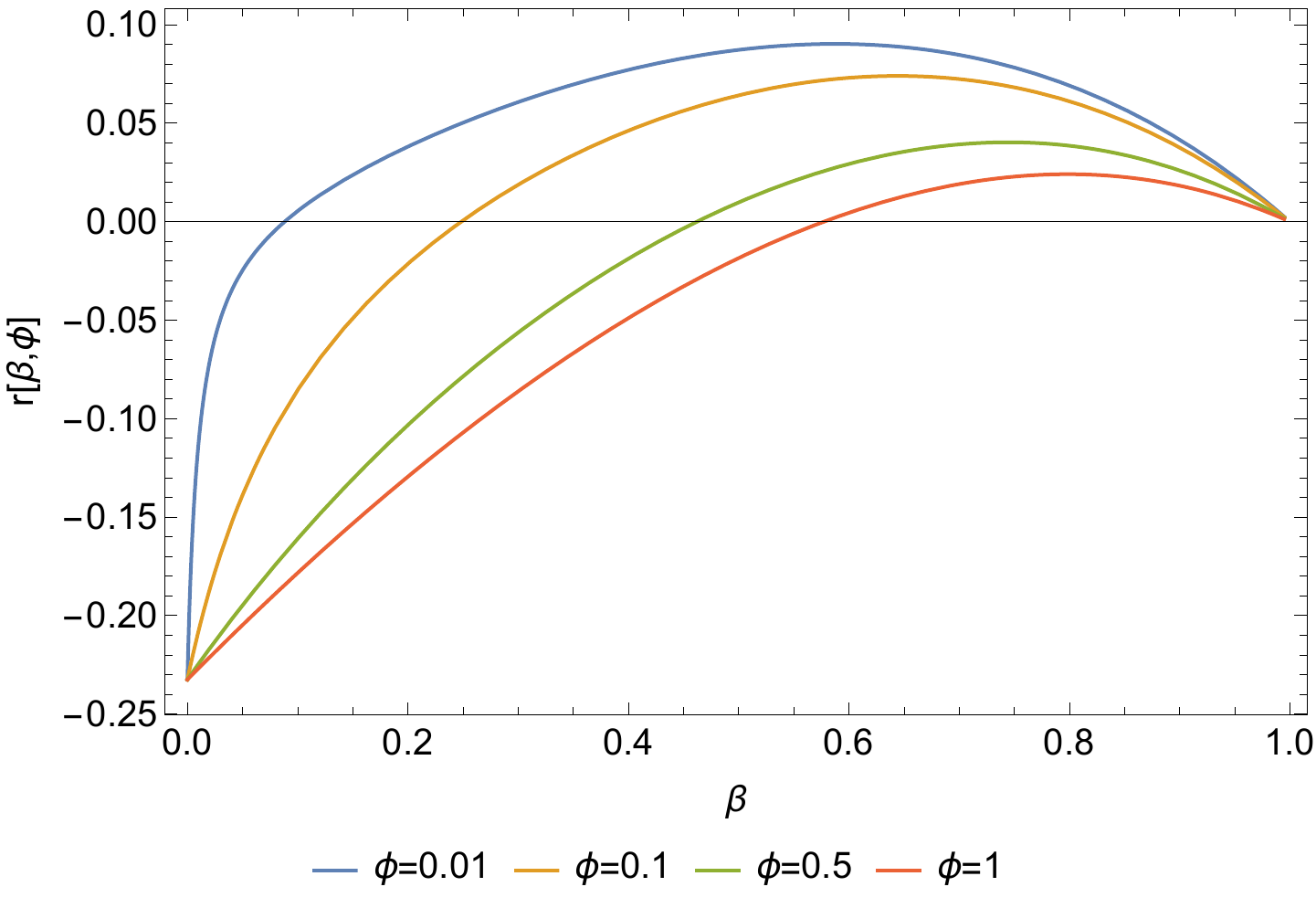}\quad
\includegraphics[width=0.45\textwidth,valign=t]{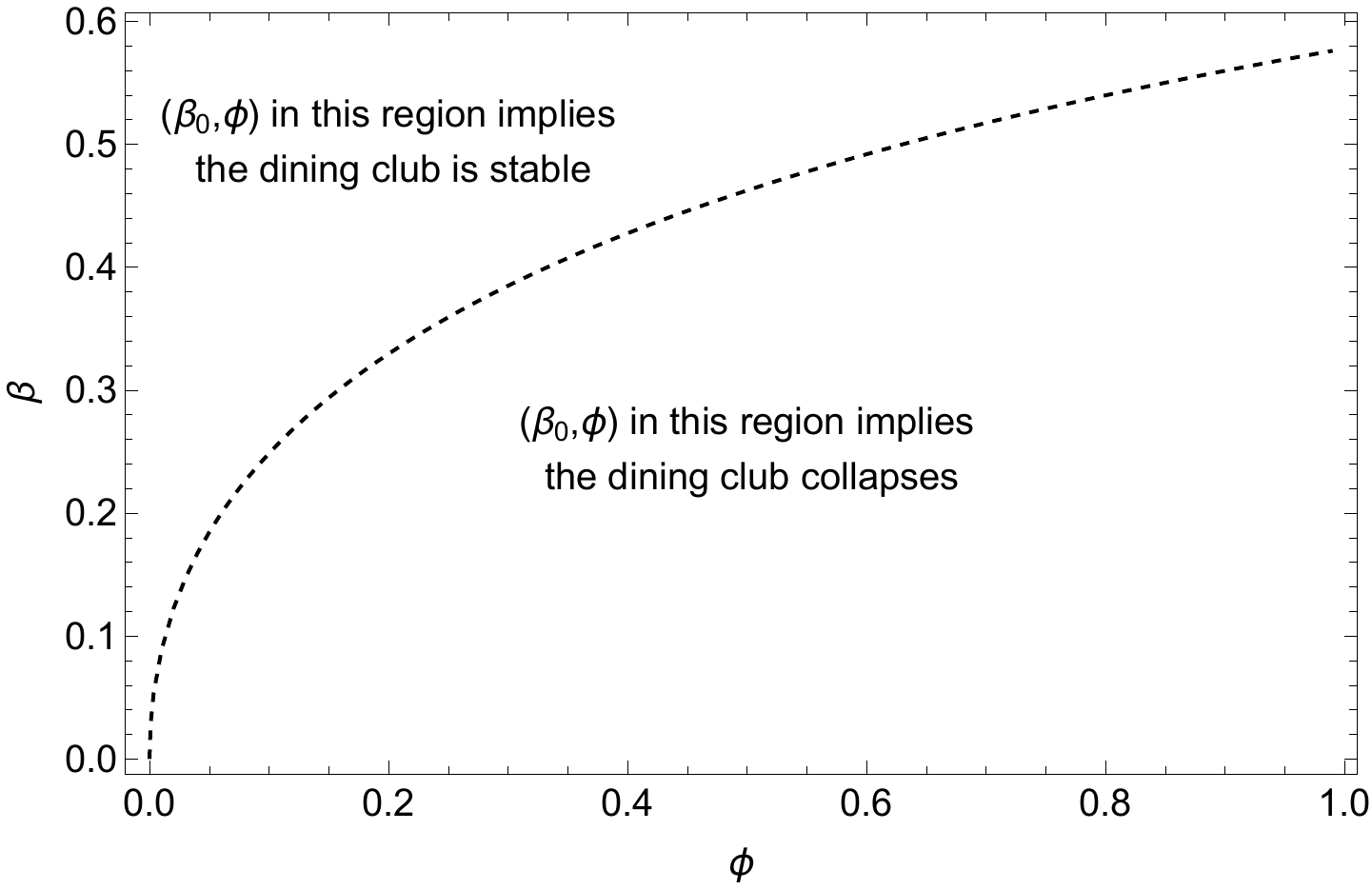}
\caption{(Left) The rate function $r(\beta,\phi)$ for varying values of $\phi$ shows that $r(t)$ changes sign as a function of $\beta$. (Right) The solution curve for $\beta^*$ as a function of $\phi$ so that $r(\beta^*,\phi) = 0$.}
\label{fig:RateBetaPhi}
\end{figure}

\begin{figure}[htbp]
\centering
\includegraphics[width=0.5\textwidth]{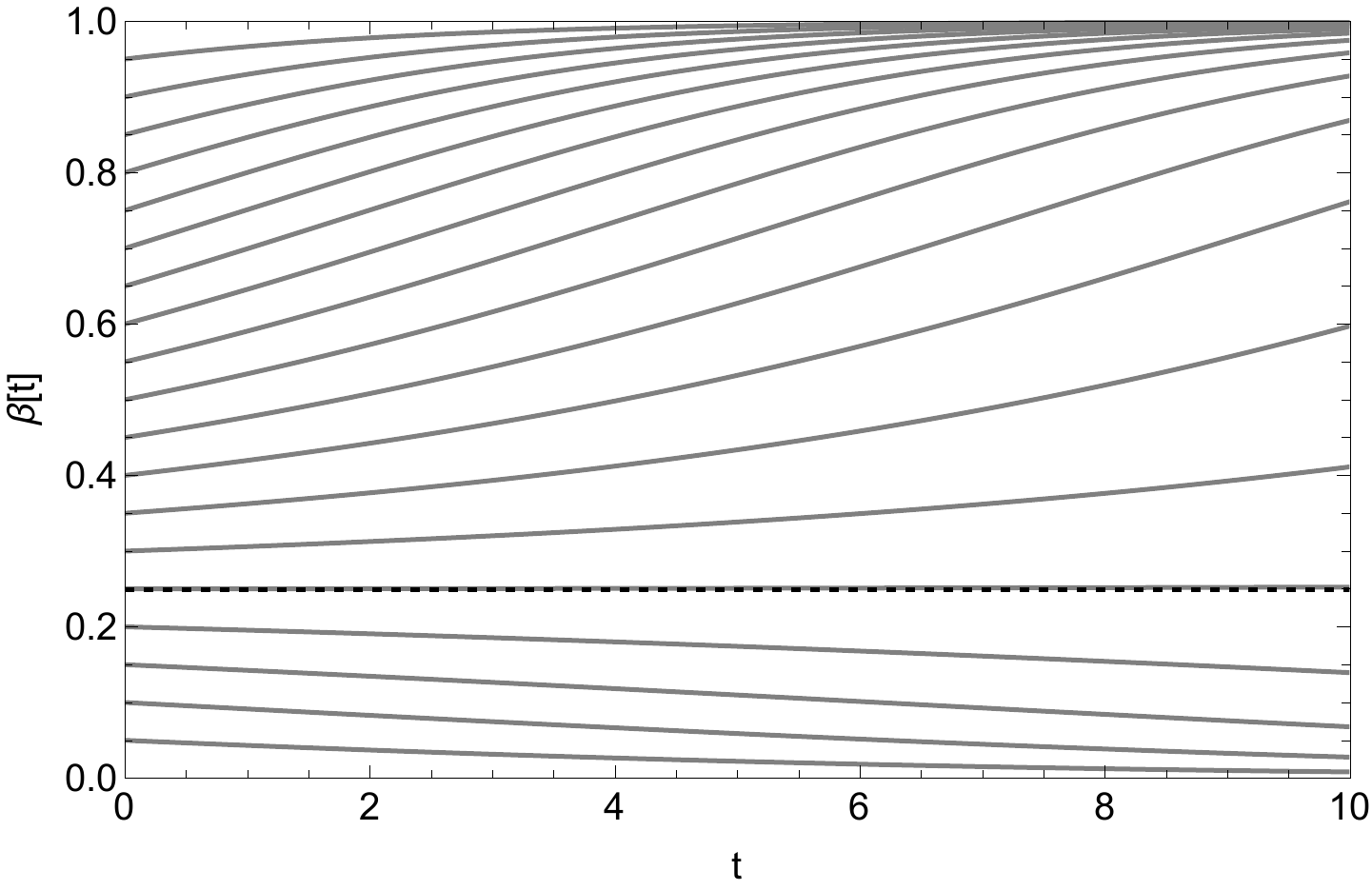}
\caption{Here $\phi = 0.1$ and we show the instability of the interior fixed point. With $\beta(0) > \beta^*$ all members of the population are eventually driven to join the dining club. If $\beta(0) < \beta^*$, the dining club fails as a result of freeloading.}
\label{fig:ExampleFlow}
\end{figure}

Suppose a proportion $\phi \in [0,1]$ of the independent population \textit{that does not eat} can deceptively pose as club members, thereby sharing in the communally available food. In the presence of a food tax, the resulting decision to join the dining club now becomes a public goods problem. Then the expected meal size to anyone receiving shared food is given by
\begin{equation*}
\tilde{p}_g(\beta) = \frac{\kappa gp_g(\beta)}{n[1-p_n(\beta)]\phi + [1-p_g(\beta)]g} = \frac{\alpha \kappa p_g(\beta)}{\phi [1-p_n(\beta)] + \alpha[1-p_g(\beta)]},
\end{equation*}
where $\alpha$ is defined in terms of $\beta$ in \cref{eqn:alpha}. Let $S_n$ be the random variable denoting the expected meal size for an independent member of the population. Then as a function of $\kappa$ and $\phi$,
\begin{align}
\mean{S_g} &= (1-\kappa ) p_g(\beta)+ [1-p_g(\beta)]\frac{\alpha  \kappa  p_g(\beta) }{\alpha  [1-p_g(\beta)]+ [1-p_n(\beta)] \phi }\; \text{and} \label{eqn:SgFreeloader}\\
\mean{S_n} &= p_n(\beta) + [1-p_n(\beta)]\phi\frac{\alpha  \kappa  p_g(\beta)  }{\alpha  [1-p_g(\beta)] +  [1-p_n(\beta)]\phi }.
\label{eqn:SnFreeloader}
\end{align}
It is possible but unwieldy to compute $r(\beta,\phi) = \mean{S_g} - \mean{S}$ using the expected meal size with deception rate $\phi$ and group size $\beta$. Plotting sample curves for $r(\beta,\phi)$ shows that the growth rate now changes sign at some value $\beta(\phi)$; see \cref{fig:RateBetaPhi} (left).
As a consequence of this, the replicator equation for $\beta$ is given by
\begin{equation*}
\dot{\beta} = \beta\left(\mean{S_g} - \mean{S}\right). 
\end{equation*}
These dynamics exhibit a new unstable equilibrium point, illustrating a bifurcation in parameter $\phi$ with numerically computed bifurcation diagram shown in \cref{fig:RateBetaPhi} (right). An example solution flow (for various initial conditions) is shown in \cref{fig:ExampleFlow}.
We can compute   $\beta^* \approx 0.577$ for $\phi = 1$. This is particularly interesting because we have essentially constructed a public goods problem in which joining the dining club enforces a taxation rate of $\kappa = 1 - p_g(\beta)$ on the members, who are then guaranteed (the public good of) a meal each day. The presence of freeloaders destabilizes the group formation process, but does not guarantee that a group cannot form. Since $\beta^*(\phi)$ is monotonically increasing, it follows that if $\phi$ grows slowly enough so that at any time $\beta(t) > \beta^*[\phi(t)]$, then the dining club will grow to include the entire population. If $\beta(t) < \beta^*[\phi(t)]$, then the dining club collapses. We impose an evolutionary dynamic on the freeloaders in the next section to study this effect. 

\section{Evolving Freeloaders}\label{sec:EvolveCheating}
If we divide the population into three groups, dining club members ($g$), non-dining club freeloaders ($f$) and non-dining club non-freeloaders ($h$), we can construct an evolutionary dynamic for the freeloaders. Let $\chi$ be the proportion of the population that is not in the dining club and will freeload (cheating) and $\eta = 1 - \beta - \chi$ to be the proportion of the population that is not in the dining club and not freeloading (honest). Then the population of freeloaders is $\chi(n + \alpha n)$. The expected meal size to any agent accepting communal food is then
\begin{multline}
\frac{\kappa gp_g(\beta)}{g[1-p_g(\beta)] + [1-p_n(\beta)]\chi(n+\alpha n)} = 
\frac{\kappa \alpha p_g(\beta)}{\alpha [1-p_g(\beta)] + [1-p_n(\beta)]\chi(1+\alpha)} =\\ \frac{\kappa \alpha p_g(\beta)}{\alpha [1-p_g(\beta)] + [1-p_n(\beta)]\chi(1-\beta)^{-1}}.
\label{eqn:NewDistribution}
\end{multline}

Let $S_g$ be as before, and let $S_f$ be the random variable denoting the meal size for an individual in the freeloading group and $S_h$ be the random variable denoting meal size for an individual from the non-freeloading non-dining club group. It follows from \cref{eqn:SgFreeloader,eqn:SnFreeloader,eqn:NewDistribution} that
\begin{align*}
&\mean{S_g} = (1-\kappa ) p_g(\beta)+ [1-p_g(\beta)]\frac{\alpha  \kappa  p_g(\beta) }{\alpha  [1-p_g(\beta)]+ [1-p_n(\beta)] \chi(1-\beta)^{-1} },\\
&\mean{S_f} = p_n(\beta) + [1-p_n(\beta)]\frac{\alpha  \kappa  p_g(\beta)  }{\alpha  [1-p_g(\beta)] +  [1-p_n(\beta)]\chi(1-\beta)^{-1} }, \text{ and}\\
&\mean{S_h} = p_n(\beta).
\end{align*}
Here, we have replaced $\phi$ with its definition in terms of $\chi$ and $\beta$. Employing the same reasoning we used to obtain \cref{eqn:BetaReplicator}, we can construct replicator equations for proportions $\beta$, $\chi$ and $\eta$.

\begin{figure}[htbp]
\centering
\includegraphics[width=0.5\textwidth]{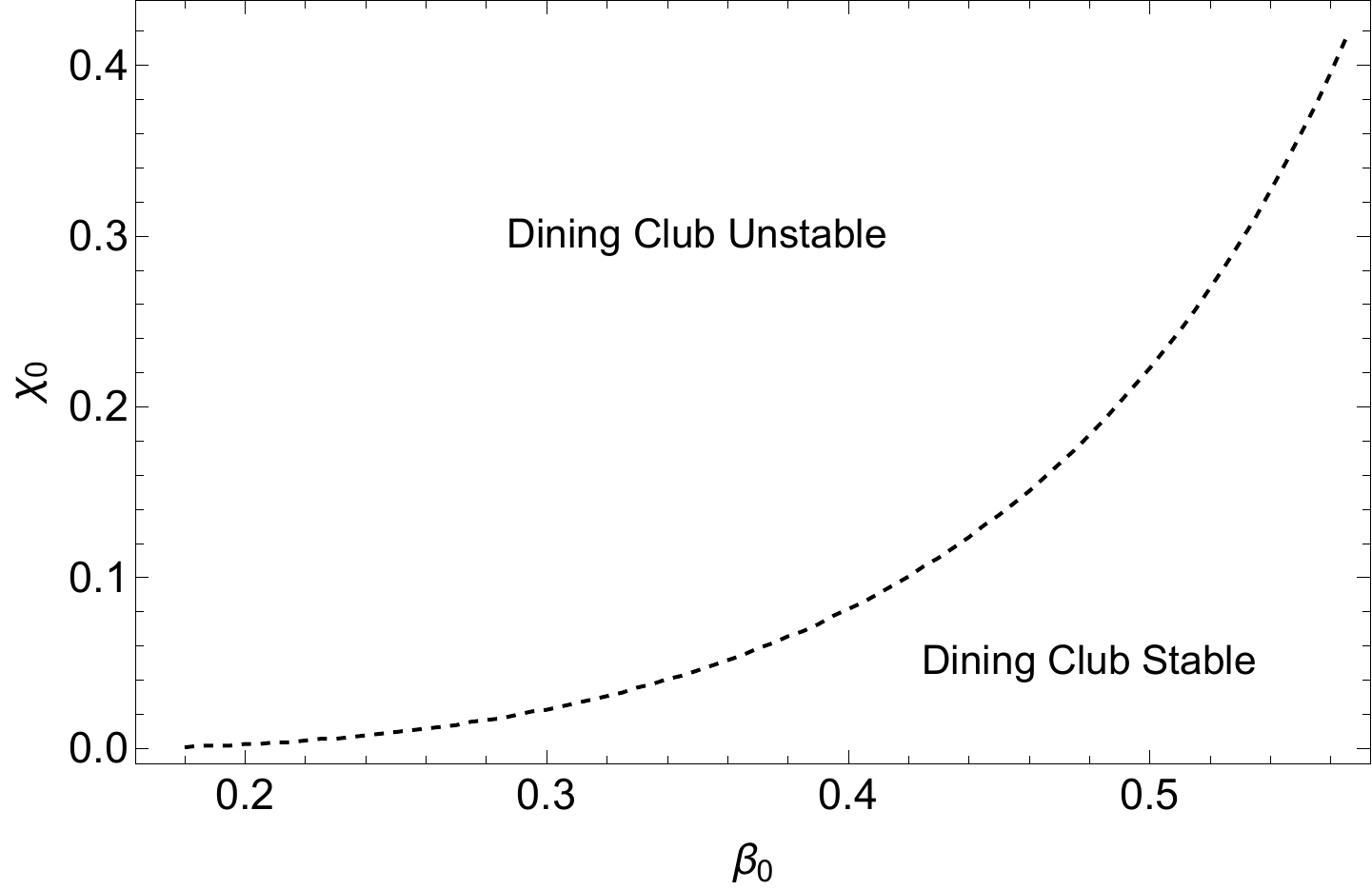}
\caption{(Right) Numerically computed curve showing the boundary between the stable and unstable dining club strategy for varying initial conditions.}
\label{fig:InitConditions}
\end{figure}

The population mean meal size is 
\begin{equation*}
\mean{S} = \chi\mean{S_f} +\beta\mean{S_g} +  \eta\mean{S_h}.
\end{equation*}
The dynamics of $\eta$ (the non-freeloading, non-dining club group) are extraneous, and we can focus on the two-dimensional system
\begin{align*}
&\dot{\beta} = \beta\left(\mean{S_g} - \mean{S}\right)\\
&\dot{\chi} = \chi\left(\mean{S_f} - \mean{S}\right),
\end{align*}
which do not depend on the value of $\eta$. 
\begin{figure}[htbp]
\centering
\includegraphics[width=0.45\textwidth]
{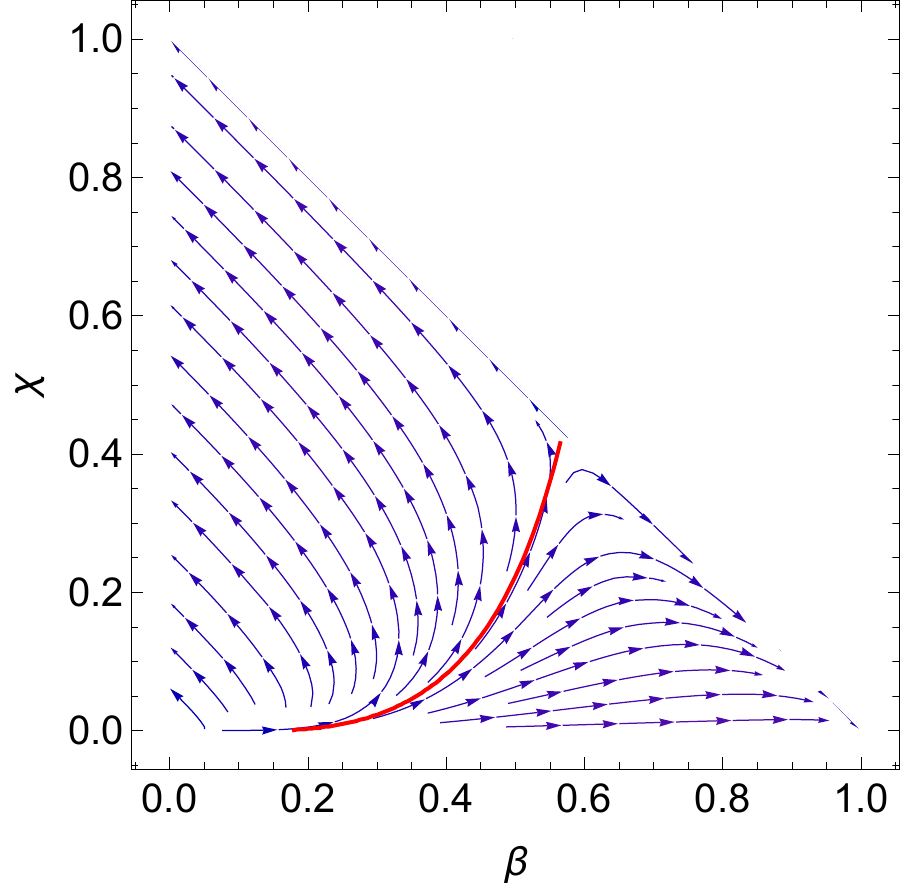}
\caption{A phase portrait of the two-dimensional system showing the dynamics of $(\beta,\chi)$. The red curve shows a numerically computed boundary between the basin of attraction of $(\beta,\chi) = (1,0)$ and $(\beta,\chi) = (0,1)$.}
\label{fig:PhasePortrait}
\end{figure}
\cref{fig:PhasePortrait} shows the dynamics of this evolutionary system. It is straightforward to compute that when $\beta = 0$, then $\mean{S_g} - \mean{S} = \mean{S_f} - \mean{S} = 0$ for all values of $\chi \in [0,1]$.  Thus, the dynamics freeze on the left boundary of the simplex 
\begin{equation*}
\Delta_2 = \left\{(\beta, \chi) \in \mathbb{R}^2 : \beta + \chi \leq 1,\;\beta \geq 0,\;\chi \geq 0\right\}.
\end{equation*}
There is a single hyperbolic saddle on the boundary of $\Delta_2$ that can be numerically computed as $(\beta,\chi) \approx (0.578, 0.422)$. The two boundary equilibria $(\beta,\chi) = (1,0)$ and $(\beta,\chi) = (0,1)$ are both locally asymptotically stable. Thus, the long-run population behaviour is dependent on the initial conditions. We can numerically construct a curve of initial conditions showing this dichotomous behaviour. This is shown in \cref{fig:InitConditions} and as the red curve in \cref{fig:PhasePortrait}.
As $\beta_0$ approaches $\beta^* \approx 0.578$ corresponding to equilibrium point for $\phi = 1$, the curve stops because $\chi_0$ would need to lie outside the simplex to cause the dining club to collapse. It is interesting to note that the phase portrait illustrates trajectories in which both $\beta$ and $\chi$ are increasing up to a point, followed by either the collapse of the dining club (while $\chi$ continues to increase) or the collapse of the freeloading group, as all population members join the dining club (and $\beta$ continues to increase).

\section{Numerical Results on Multiple Dining Clubs}\label{sec:MultipleClubs}
We now consider KPRP with two dining clubs. We model three groups of agents $\mathcal{F}$, $\mathcal{G}_1$ and $\mathcal{G}_2$ for free agents, dining club one and dining club two respectively. We estimate $\mean{S_{g_1}}$, $\mean{S_{g_2}}$ and $\mean{S_f}$  using Monte Carlo simulation. This Monte Carlo simulation is then embedded into a larger dynamic process for updating the groups.

In the Monte Carlo simulation, the free agent group acts normally, choosing a restaurant randomly. The members of the dining clubs also chose restaurants randomly, but with the constraint that no two agents in a dining club may choose the same restaurant. Since we are studying this system numerically, we introduce two kinds of taxation policies.
\begin{enumerate}
    \item Policy I:  We assume a given tax rate $\kappa$ with no redistribution; i.e., the tax goes to maintain the dining club in some form.
    \item Policy II: Agents within the dining club are taxed at a rate $\kappa$ given by, \cref{eqn:OptimalKappa} and food is redistributed to club  members who do not eat (and possibly freeloaders).
\end{enumerate}

Agents in the free market will randomly choose a dining club to eat in if they do not get food on a given day with probability $1$. That is, we assume $\phi = 1$. We also introduce a probability $\rho$ that cheaters will be caught. If a cheater gets caught, their food is not distributed and becomes waste.

In the dynamic model that follows, we refer to the process of simulating groups eating over several days by the function $\texttt{MonteCarlo}(\mathcal{F},\mathcal{G}_1,\mathcal{G}_2, \kappa, \rho)$. The system dynamics of our simulation are then described by the following steps:
\begin{algorithmic}[1]
\STATE \underline{\textbf{Input:}} $\mathcal{F}$, $\mathcal{G}_1$, $\mathcal{G}_2$.
\WHILE{There is at least one agent in each group}
\STATE Compute $(\mean{S_{g_1}}, \mean{S_{g_2}}, \mean{S_f}) = \texttt{MonteCarlo}(\mathcal{F},\mathcal{G}_1,\mathcal{G}_2,\kappa,\phi)$.
\STATE Set $\mathcal{P} = \mathcal{F} \cup \mathcal{G}_1 \cup \mathcal{G}_2$.
\STATE Choose two agents $i$ and $j$ at random from $\mathcal{P}$.
\STATE Let $\texttt{Group}(i)$ (resp. $\texttt{Group}(j)$) be the group to which $i$ (resp. $j$) belongs.
\STATE Let $p_i$ (resp. $p_j$) be the probability that $i$ (resp. $j$) eats.
\IF{$p_i > p_j$}
\STATE Move $j$ to $\texttt{Group}(i)$
\ELSIF{$p_j > p_i$}
\STATE Move $i$ to $\texttt{Group}(j)$
\ENDIF
\STATE Remove $i$ and $j$ from $\mathcal{P}$.
\IF {$|\mathcal{P}| > 1$}
\STATE \textbf{goto} 5
\ELSE
\STATE \textbf{goto} 3
\ENDIF
\ENDWHILE
\end{algorithmic}
It is clear in the dynamics simulated by this model, there are three equilibria corresponding to the cases when all agents are in $\mathcal{F}$ or $\mathcal{G}_1$ or $\mathcal{G}_2$.

\subsection{Simulation Results}
For each simulation, we divide 100 agents into $\mathcal{F}$, $\mathcal{G}_1$ and $\mathcal{G}_2$. To construct an approximation for the basins of attraction for three equilibrium populations, we ran the simulation using 1000 replications simulation and every possible (discrete) starting condition on $|\mathcal{F}|$, $|\mathcal{G}_1|$ and $|\mathcal{G}_2|$. 

\begin{figure}[htbp]
\centering
\subfloat[$\kappa = 0.05$]{\includegraphics[width=0.42\textwidth]{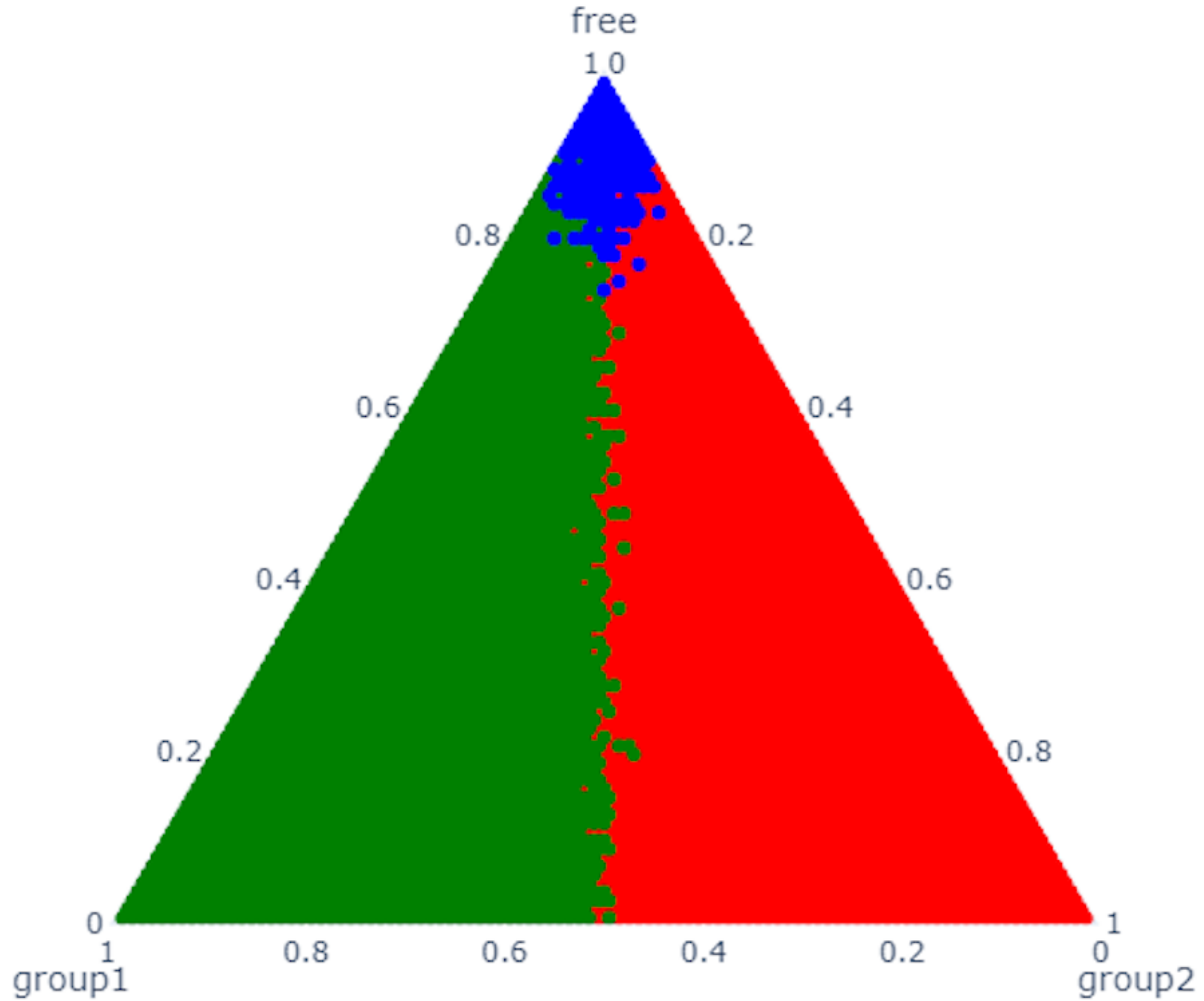}} 
\subfloat[$\kappa = 0.07$]{\includegraphics[width=0.42\textwidth]{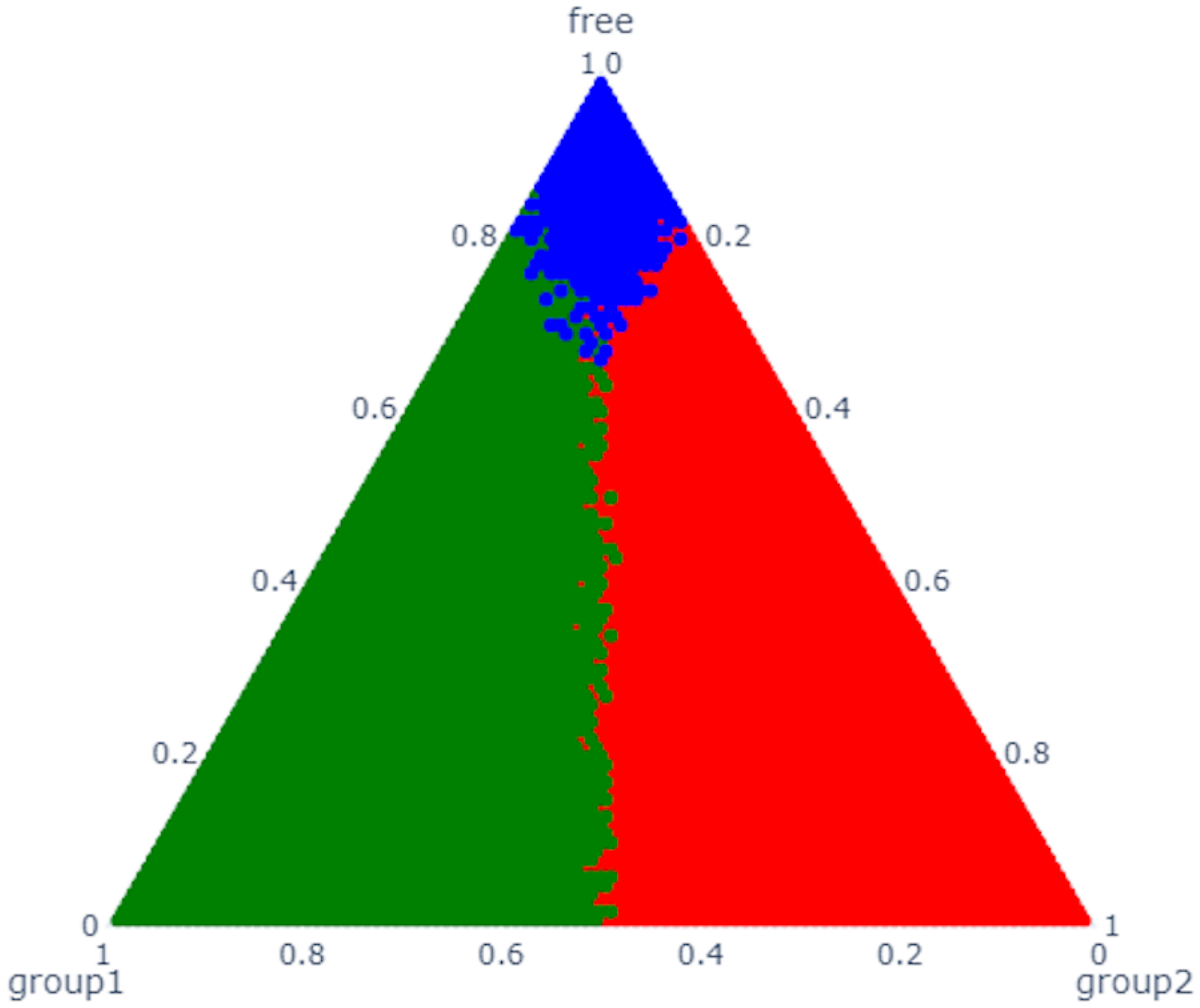}}\\
\subfloat[$\kappa = 0.1$]{\includegraphics[width=0.42\textwidth]{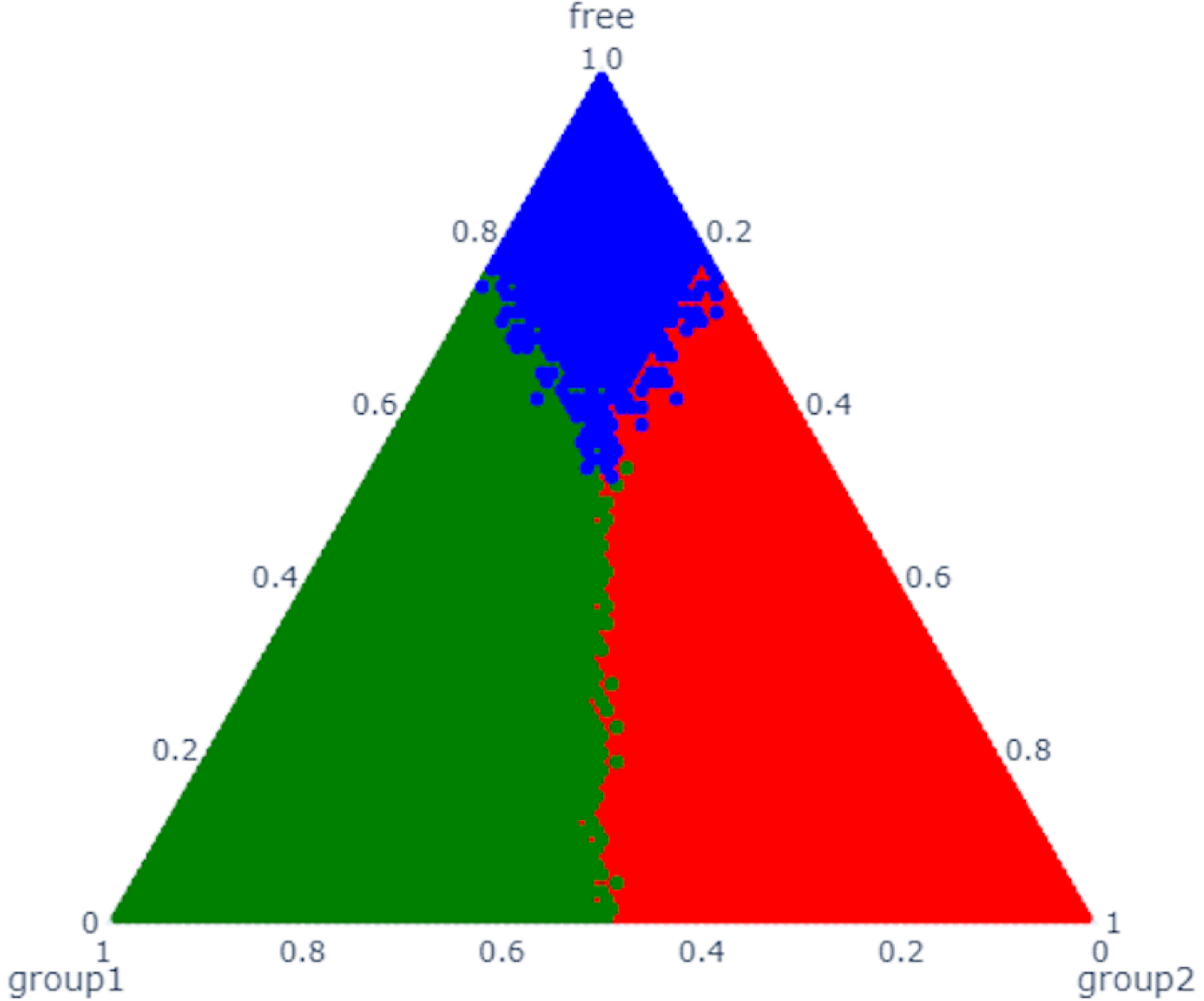}}
\subfloat[$\kappa = 0.15$]{\includegraphics[width=0.42\textwidth]{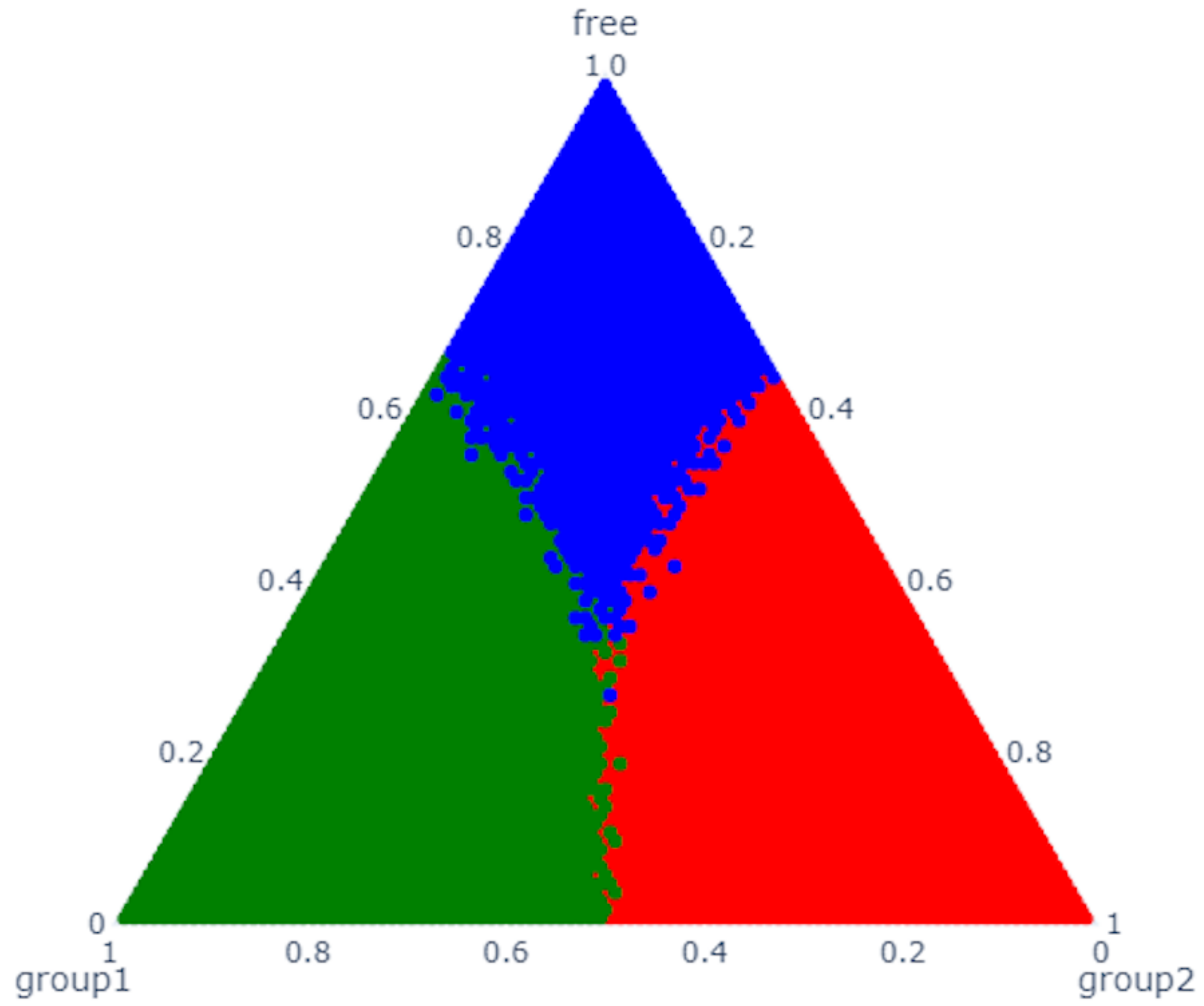}}\\
\subfloat[$\kappa = 0.2$]{
\includegraphics[width=0.45\textwidth]{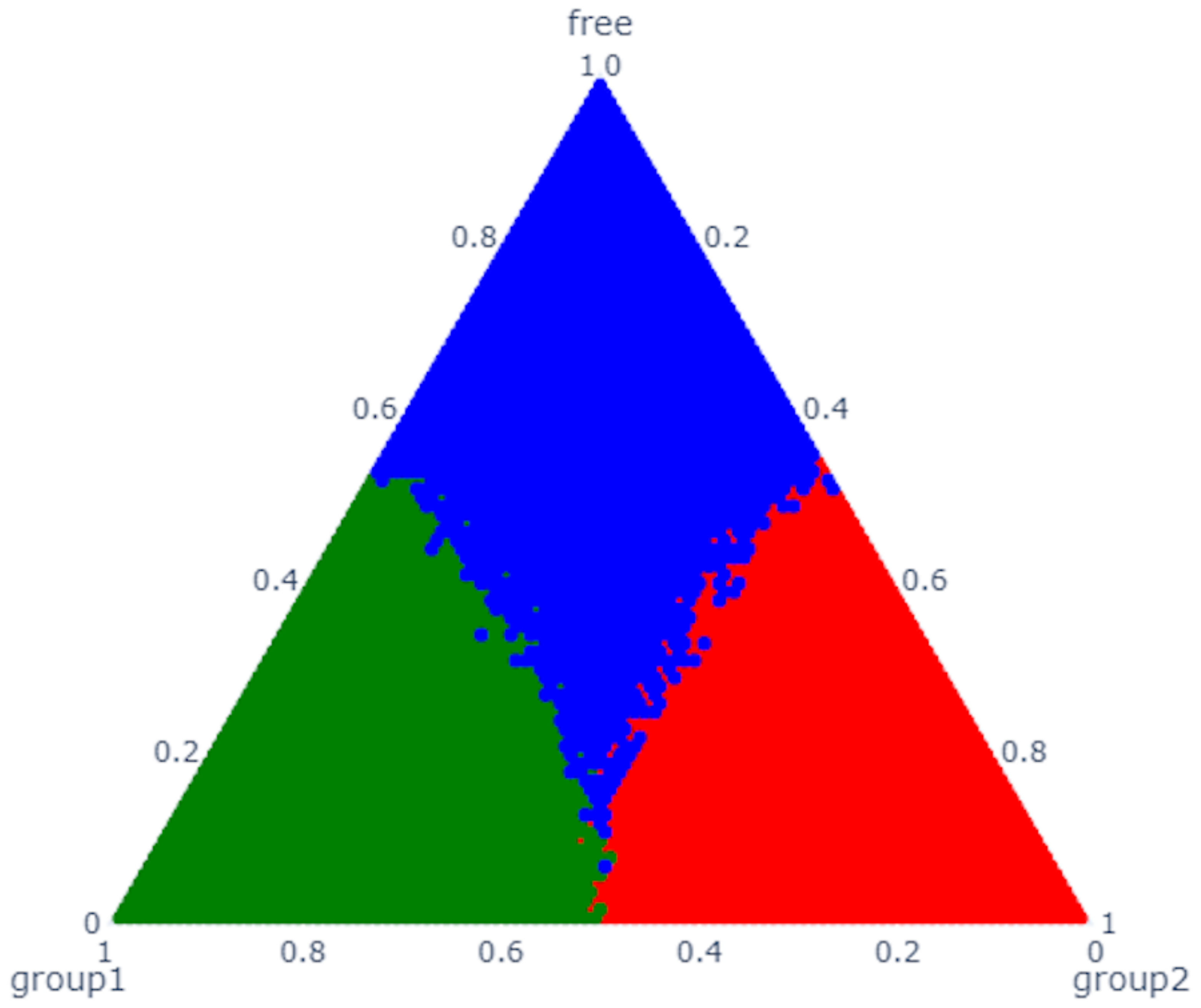}
}\\
\includegraphics[width=0.55\textwidth]{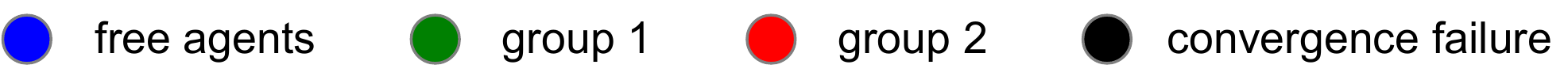}
\caption{Basins of attraction for various tax rates are shown in a ternary plot. The different colours indicate where the model converges from the given starting point. 
}
\label{fig:Policy1}
\end{figure}

\paragraph{Tax Policy I:} We explore the effect of varying $\kappa$ from $0.05$ to $0.15$. To manage simulation time, we executed the while loop at most, 10000 times. If all players had not joined a single community by then, we declared this a failed run, suggesting slow convergence from this initial condition. The outcome of almost all experiments resulted in a dominant group (either free agents or dinning clubs) being formed. This is illustrated in \cref{fig:Policy1}.
Let $\beta_1$ and $\beta_2$ be the proportion of the population in dining clubs one and two, respectively, and let $\nu = 1 - \beta_1 - \beta_2$ be the free group proportion. Then the dynamics can be projected to the two-dimensional unit simplex $\Delta_2$ embedded in $\mathbb{R}^3$ with coordinates $(\beta_1,\beta_2,\eta)$. When the simulation converges, can determine the $\omega$-limit set of trajectories leaving (near) an initial condition $(\beta_1^0,\beta_2^0,\eta^0)$. \cref{fig:Policy1} shows that the size of the tax rate $\kappa$ is correlated with the size of the basin of attraction for the free agent group. The dynamics roughly partition the simplex into three basins of attraction, with the basins of attraction for the two dining clubs exhibiting symmetry as expected. On the boundaries of these regions, we expect unstable coexistence of multiple groups would be possible. This is qualitatively similar to the unstable fixed point identified in \cref{fig:ExampleFlow}.

\begin{figure}[htbp]
\centering
\includegraphics[width=0.45\textwidth]{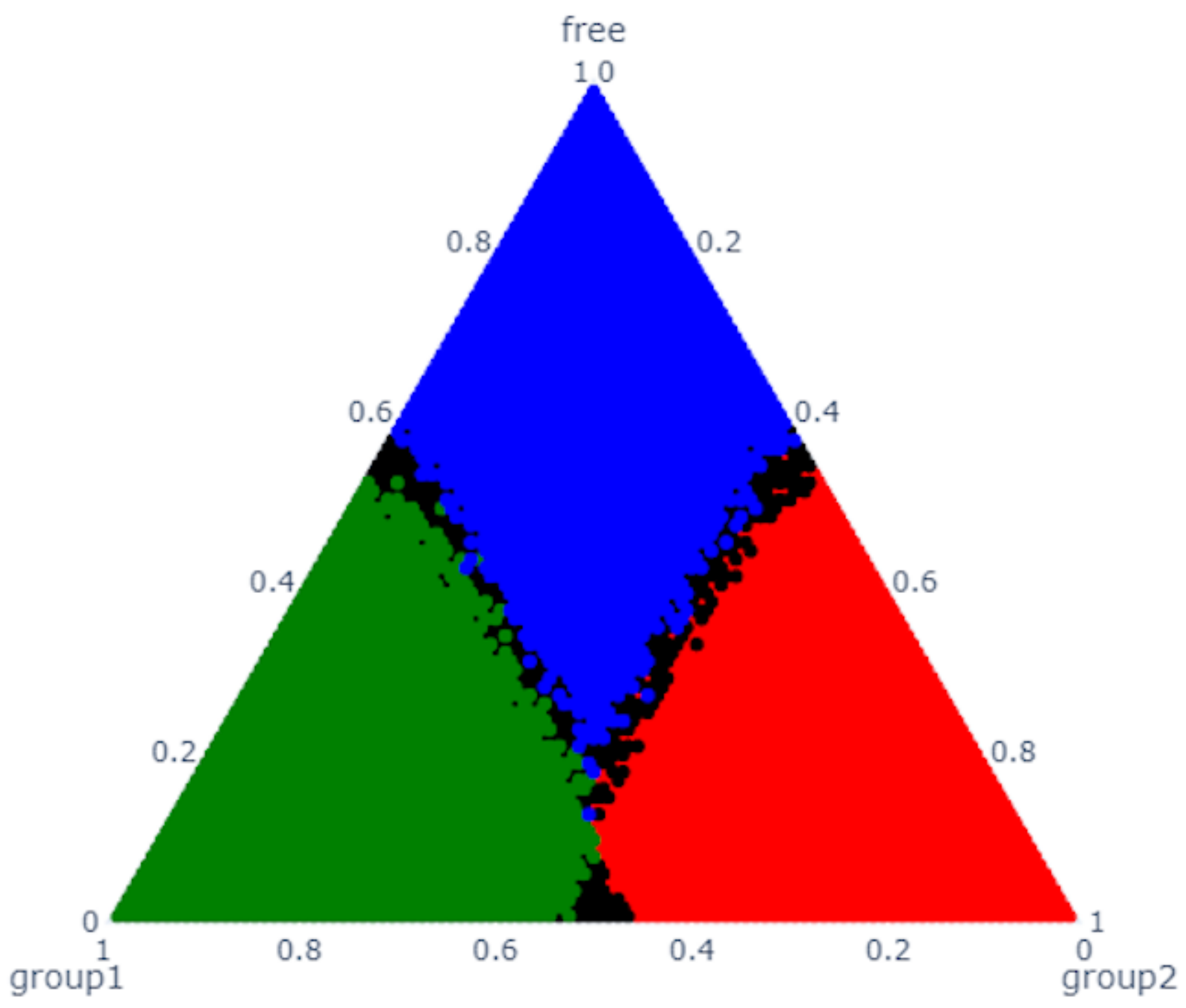} 
\includegraphics[width=0.45\textwidth]{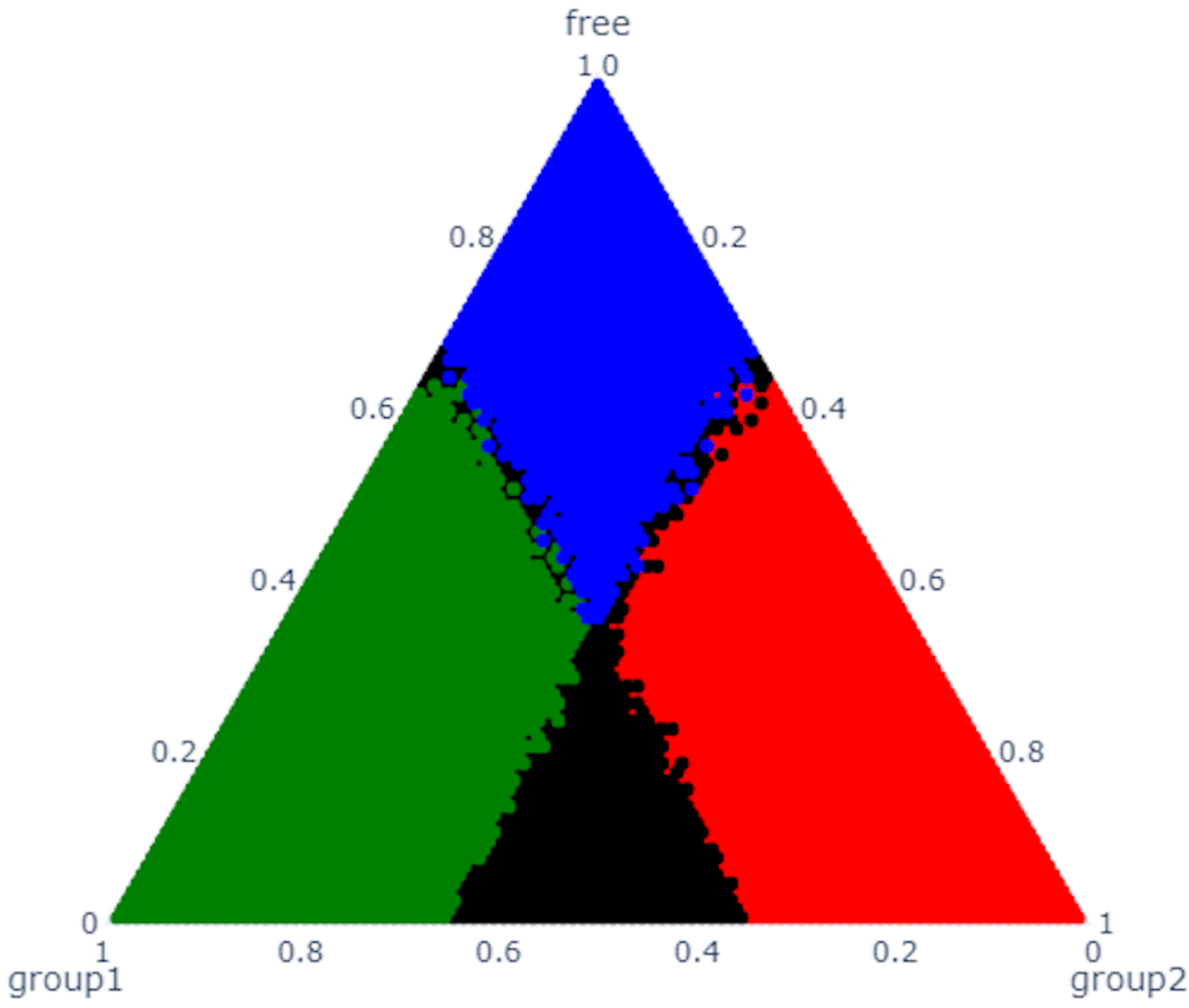}\\
\includegraphics[width=0.4\textwidth]{Figures/Legend.pdf}
\caption{(Left) We show the basins of attraction when the probability that a cheater is caught is set at $0.5$. (Right) Basins of attraction when the probability that a cheater is caught is $1$.}
\label{fig:Policy2}
\end{figure}

\paragraph{Tax Policy II:}
In a second set of experiments, we let $\rho$ vary between $0$ and $1$ and used \cref{eqn:OptimalKappa} to set the tax policy. The cheating probability was fixed at $\phi = 1$. As before, we executed the while loop at most, 10000 times. If all players had not joined a single community by then, we declared this a failed run, suggesting slow convergence from this initial condition. Basins of attraction for various fixed points are shown in \cref{fig:Policy2}.
It is interesting to notice that there are a substantial number of failed cases between the clubs. This suggests an area of slow dynamics and possibly the existence of a slow manifold. Constructing a mathematical model of this scenario is an area reserved for future work, since it is unclear exactly how the dynamics are changing in this region.

\section{Conclusions and Future Directions}\label{sec:Conclusions}
In this paper, we studied the Kolkata Paise Restaurant Problem (KPRP) with dining clubs. Agents in a dining club mutually agree to visit separate restaurants, thereby increasing the probability that they eat (obtain a resource). An evolutionary game model was formulated describing the choice to join the dining club. We showed that joining the dining club is an evolutionarily stable strategy, even when members are taxed (in food) and resources are distributed. When cheating was introduced to the non-dining club members, i.e. the non-dining club members could deceptively benefit from the communal food collected by the dining club, a new unstable fixed point appears. We analysed this bifurcation as well as the decision to cheat using the resulting replicator dynamic. Numeric experiments on two dining clubs show that the behaviour in this case is similar to the case with one dining club, but may exhibit richer dynamics. 

There are several directions for future research. Studying the theoretical properties of two (or more) dining clubs is clearly of interest. Adding many groups (i.e., so that the number of groups is a proportion of the number of players) might lead to unexpected phenomena. Also, allowing groups to compete for membership (by varying tax rates) might create interesting dynamics. As part of this research, investigation of the dynamics on the boundary both in theory and through numeric simulation would be of interest. A final area of future research would be to investigate the effect of taxing cheaters who are caught, thus allowing them to eat, but discouraging them from cheating. Determining the impact on the basins of attraction in this case would be the primary research objective. 

\section*{Acknowledgements}
A.H., A.B., and C.G. were supported in part by the National Science Foundation under grant DMS-1814876.

\bibliographystyle{elsarticle-num} 
\bibliography{KolkotaPaise-ArXiv}

\end{document}